\documentstyle[aps,prb,floats,epsfig]{revtex}

\begin{document}
%
%\draft
%\title{Calculation of Elastic Green's Functions for Lattices with Cavities}
%\author{J. Schi{\o}tz\cite{JSaddr} and A. E. Carlsson}
%\address{Department of Physics, Washington University, St.\ Louis,
%  Missouri 63130}
%\date{\today}
%\maketitle
%
\twocolumn[%
\begin{center}
  \large\bf
  Calculation of Elastic Green's Functions for Lattices with Cavities
\end{center}
\begin{center}
  J. Schi{\o}tz\cite{JSaddr}, and A. E. Carlsson\\
  {\em Department of Physics, Washington University, St.\ Louis, MO
    63130-4899, USA}\\
  (December 17, 1996)
\end{center}
\begin{quote}
\begin{quote}\setlength{\parindent}{1em}
%\begin{abstract}
  In this Brief Report, we present an algorithm for calculating the
  elastic Lattice Greens Function of a regular lattice, in which
  defects are created by removing lattice points.  The method is
  computationally efficient, since the required matrix operations are
  on matrices that scale with the size of the defect subspace, and not
  with the size of the full lattice.  This method allows the treatment
  of force fields with multi-atom interactions.
%\end{abstract}
\end{quote}
\end{quote}
]

%\narrowtext

Lattice Green's function methods provide a very computationally efficient
way of handling the long-ranged stress fields around defects in materials.
The purpose of this Report is to generalize the lattice Green's
function method of Thomson {\em et al.\/}\cite{ThZhCaTe92} to problems
in which a set of lattice points has been completely removed from the
problem.  Such problems may be the modeling of the elastic fields in
crystalline systems in which cavities are created.
An example of such systems that have been modeled in this
way are blunt cracks\cite{ScCaCaTh96,ScCaCa97}.  This opens for the
application of Green's function techniques to a new class of problems:
The previously published method\cite{ThZhCaTe92,CaCaTh95} has been
used to study lattice defects where {\em connections\/} between
lattice points have been
altered\cite{ThZhCaTe92,ZhCaTh93,ZhCaTh94,ZhCaTh94b,ThZh94,Th95} (for
example leading broken to introduce cracks).  The present extension
of the method permits the study of problems where {\em lattice
points\/} (atoms) have been removed from the system.  
The methodology was used in the calculations of 
Refs. \onlinecite{ScCaCaTh96} and \onlinecite{ScCaCa97}, but the derivation has
not been presented in the literature yet.

In the lattice Green's function method, one begins with a
force constant matrix $\Phi_{ij}^{ab}$ defined by
\begin{equation}
  E = {1 \over 2} u_i^a \Phi_{ij}^{ab} u_j^b
\end{equation}
where $E$ is the total energy, $u_i^a$ is the displacement in the $i$
direction of atom $a$ and summation over repeated indices is implied.
$\Phi_{ij}^{ab}$ is thus the negative of the force in the $i$
direction on atom $a$ when atom $b$ is displaced a unit distance in
the direction $j$:
\begin{equation}
  \label{eq:force}
  F_i^a = - \Phi_{ij}^{ab} u_j^b
\end{equation}

For a simple pair potential with radial forces between atoms,
$\Phi_{ij}^{ab}$ is given by
\begin{mathletters}
  \label{eq:pairpot}
  \begin{eqnarray}
    \Phi_{ij}^{ab} &=& - {r_i^{ab} \over \left| {\bf r}^{ab} \right|}
    \lim_{u_i^b \rightarrow 0} {f \left( \left| {\bf r}^{ab} + {\bf
        e}_j u_j^b \right| \right) \over u_j^b }, \qquad a \neq b
    \label{eq:pairpotoffdiag} \\
    \Phi_{ij}^{aa} &=& - \sum_b {r_i^{ab} \over \left| {\bf r}^{ab}
    \right|} \lim_{u_i^a \rightarrow 0} {f \left( \left| {\bf r}^{ab}
    - {\bf e}_j u_j^a \right| \right) \over u_j^a. }
    \label{eq:pairpotdiag}
  \end{eqnarray}
\end{mathletters}
where $f(r)$ is the force between the two atoms, ${\bf r}^{ab}$
is the vector between the equilibrium positions of atoms $a$ and $b$
and ${\bf e}_i$ is a unit vector in the $i$ direction.  The
$r_i^{ab} / \left| {\bf r}^{ab} \right|$ are a projection operators.

The Greens function gives the formal solution of equation (\ref{eq:force}) 
for $u$:
\begin{eqnarray}
  \label{gfdef}
  u_i^a &=& G_{ij}^{ab} F_j^b,\\
  {\bf G} &=& {\bf \Phi}^{-1} \label{eq:definition}
\end{eqnarray}
It is inconvenient to label the degrees of freedom by two indices (the
atom and the component of the displacement). Therefore, we 
introduce a single labeling of all degrees of freedom, leading naturally
to a matrix notation where a component of ${\bf G}$, $G_{ij}$, is the
response by the degree of freedom $i$ to a force acting on the degree
of freedom $j$.

Thomson {\em et al.\/} demonstrate\cite{ThZhCaTe92} how the Greens
function of a regular lattice can be calculated in Fourier space in a
time that scales linearly with the number of lattice points ($N_{lat}$).
The Fourier-space approach breaks down when defects are present, breaking
translational symmetry. Thomson {\em et al.\/} showed
that defects can be treated without a complete "brute-force" inversion
of equation (\ref{eq:definition}), by the use of multiple-scattering theory.
If the values of only a small
number of ``bonds'' (i.e.\ elements in the $\Phi$ matrix) are changed,
the new Greens function ${\bf G}$ can be calculated from the
perfect-lattice Greens function ${\bf G}^0$ using a Dyson equation:
\begin{equation}
  \label{eq:dyson}
  {\bf G} = {\bf G}^0 \bigl( {\bf 1} - {\bf \delta \Phi} {\bf G}^0
\bigr)^{-1}
\end{equation}
where ${\bf \delta \Phi}$ is the change in the force constant matrix.
This calculation scales as $N_{def}^3$, where $N_{def} \ll
N_{lattice}$ is the number of rows or columns in ${\bf \delta \Phi}$
containing non-zero elements, i.e.\ the number of degrees of freedom
of atoms where one or more bonds are modified, and $N_{lattice}$ is
the total number of degrees of freedom in the lattice.

%%%%%%%%%%%%%%%%%%%%
%%%%            %%%%
%%%%  FIGURE 1  %%%%
%%%%            %%%%
%%%%%%%%%%%%%%%%%%%%
\begin{figure}
  \begin{center}
    \leavevmode
    \epsfig{file=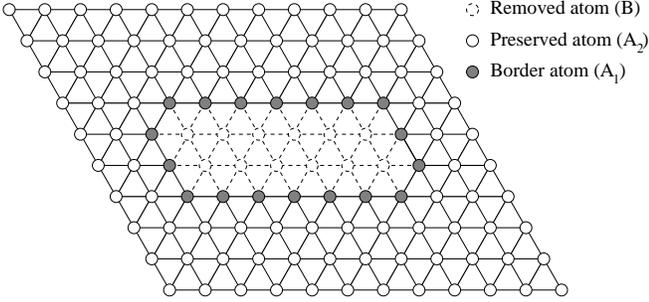,angle=-90,width=\linewidth}
  \end{center}
  \caption{A hexagonal lattice, where 12 atoms have been removed.  The
    removed atoms and their bonds are shown with dashed lines.  Border
    atoms that have lost one or more neighbors are shaded.  The
    corresponding diagonal elements of ${\bf \delta\Phi}$ are
    non-zero, since the on-site force constant has changed: when such
    an atom is displaced it ``feels'' a different force since there is
    no force from the missing neighbors.  In the text the following
    classification scheme is used: The removed atoms are in class $B$,
    the remaining atoms are in class $A$.  Class $A$ is further
    subdivided into two classes: the grey atoms, having lost one or
    more neighbors, are in class $A_1$, and the white atoms are in
    class $A_2$.}
  \label{fig:lattice}
\end{figure}
When atoms are removed from the system (see figure~\ref{fig:lattice}),
it is important to be able to calculate the Green's function as
efficiently as when only bonds are changed, i.e.\ the Greens function
should be calculated starting from the perfect-lattice Greens function
${\bf G}^0$, and not directly from equation (\ref{eq:definition}).  Let us
divide the degrees of freedom in two classes: $A$ are the degrees of
freedom of atoms that are ``kept'', and $B$ are those that are removed
from the system.  ${\bf M}_{AA}$ is then the sub-matrix of ${\bf M}$
obtained by keeping only columns and rows corresponding to degrees of
freedom in class $A$.  In a similar ways the other sub-matrices ${\bf
  M}_{BB}$, ${\bf M}_{AB}$, and ${\bf M}_{BA}$ are defined.  Further,
let ${\bf G^*}$ be the Green's function of the new system.  ${\bf
  G^*}$ cannot be calculated by breaking all bonds to the atoms to be
removed, and then applying the Dyson equation (\ref{eq:dyson}), since
the resulting Greens function is singular\cite{endnote1}.

The force-constant matrix entering ${\bf G^*}$ is perturbed relative
to the full force-constant matrix in two ways. First, it has smaller
dimension in the sense that it only contains atoms in the $A$-region.
Second, the on-site force constants (corresponding to the force on an
atom resulting from its own motion) of the $A$-region border atoms
near the cavity are changed, since these are determined by the
neighboring atoms (terms in the sum in equation (\ref{eq:pairpotdiag})
are missing).  For example, if an atom has no neighbors, then it has a
vanishing on-site force constant. Then ${\bf G^*}$ is given by
\begin{equation}
  \label{eq:gstardef}
  {\bf G^*} = \left( {\bf \Phi}_{AA} + {\bf \delta\Phi}_{AA}
     \right)^{-1},
\end{equation}
where ${\bf \delta\Phi}_{AA}$ corresponds to the change in the on-site
force constants of the border atoms.  For clarity, we define the
auxiliary Green's functions ${\bf G}^0 = {\bf \Phi}^{-1}$ for the
perfect crystal, and ${\bf G'} = ({\bf \Phi}_{AA})^{-1}$ for the
crystal with atoms removed but no changes in the force constants of
the remaining atoms.  (Note that ${\bf G'} = ({\bf \Phi}_{AA})^{-1}
\neq ({\bf \Phi}^{-1})_{AA} = {\bf G}^0_{AA}$).  The calculation of
${\bf G^*}$ then goes in two steps.  We first calculate ${\bf G'}$, as
an intermediate step:
\begin{eqnarray}
  \label{eq:Gprime}
  {\bf G'} = ({\bf \Phi}_{AA})^{-1} &=& \biggl(\Bigl(\bigl({\bf
    G}^0\bigr)^{-1}\Bigr)_{AA}\biggr)^{-1} \\
  &=& {\bf G}^0_{AA} - {\bf G}^0_{AB}
  \bigl( {\bf G}^0_{BB} \bigr)^{-1} {\bf G}^0_{BA} \nonumber
\end{eqnarray}
(The last equality is proven in the appendix).  We then calculate 
${\bf G^*}$ with a Dyson equation:
\begin{equation}
  {\bf G^*} \bigl( {\bf \Phi}_{AA} + {\bf \delta\Phi}_{AA} \bigr) =
  {\bf 1}
\end{equation}
which gives
\begin{equation}\label{eq:newdyson}
  {\bf G^*} = {\bf G'} \bigl( {\bf 1} -  {\bf G'} {\bf \delta
    \Phi_{AA}}\bigr)^{-1}.
\end{equation}

Equations (\ref{eq:Gprime}) and (\ref{eq:newdyson}) are the central
results of this Report. We note that they can be used to describe
systems with many-atom interactions such as the Embedded Atom
Method\cite{DaBa84}, the Effective Medium Theory\cite{JaNoPu87}, the
glue model\cite{ErToPa86} etc.  In these methods, the force constants
between different atoms in the $A$ region also change because the
effective interatomic potentials are environmentally dependent. This
effect can be included straightforwardly by modification of the ${\bf
  \delta\Phi}_{AA}$ term.

Finally, let us show that this method is computationally feasible.
Let the degrees of freedom that are kept be separated into two classes,
$A_1$ and $A_2$, where $A_1$ is the set of degrees of freedom of atoms
where the corresponding elements of ${\bf \delta\Phi}$ is non-zero, or
where we are going to need the Greens function; the latter would include
atoms on which nonlinear forces or loading forces will eventually be placed.
Let $n(S)$ be the number of degrees of freedom of the atoms in set $S$.
Typically, $n(A_1)$ will be a few hundreds or less, whereas $A_2$
contains millions of degrees of freedom\cite{ScCaCaTh96,ScCaCa97}.
$n(B)$ is also typically a few hundreds.  Equation (\ref{eq:Gprime})
then consists of the inversion of a $n(B) \times n(B)$ matrix, and two
$n(A_1) \times n(B)$ matrix multiplication.  Since ${\bf
  \delta\Phi}_{A_2A_2} = {\bf 0}$ equation (\ref{eq:newdyson})
consists of a $n(A_1) \times n(A_1)$ matrix inversion and a similar
multiplication.  With the relatively small sizes of $A_1$ and $B$,
this is clearly computationally practical, whereas the direct matrix
inversion in equation (\ref{eq:gstardef}) is not.  See also the
discussion of the computational burden of equation (\ref{eq:newdyson})
in Thomson {\em et al.\/}\cite{ThZhCaTe92} if there are many atoms in
$A_1$ where ${\bf \delta\Phi}$ is zero.

In conclusion, we have shown how to generate Lattice Green's functions
for lattices where the defects are not limited to perturbations of a
small number of bonds, but where lattice points have been removed. 
The Green's function is generated in a computationally efficient way,
and has already been used to simulate the elastic fields near a
blunted crack\cite{ScCaCa97}.  The method is not limited to elastic
fields in crystal lattices, but to all problems where a lattice
Green's function may be useful, and where a small number of lattice
points are removed from the problem.

The authors would like to acknowledge many useful discussions with
Robb Thomson and Lilly M. Canel.  This work was supported by the
National Institute of Standards and Technology under award
60NANB4D1587 and by the Department of Energy under grant number
DE-FG02-84ER45130.

\section*{Appendix}

We here prove that
\begin{equation}
  \label{eq:tobeproven}
  \Bigl(\bigl({\bf M}^{-1}\bigr)_{AA}\Bigr)^{-1} = {\bf M}_{AA} -
  {\bf M}_{AB} \bigl({\bf M}_{BB}\bigr)^{-1} {\bf M}_{BA}
\end{equation}
Proof: Let ${\bf N} = {\bf M}^{-1}$.  We then have ${\bf MN} = {\bf
  1}$, which can be split into four parts:
\begin{mathletters}
  \begin{eqnarray}
    {\bf M}_{AA} {\bf N}_{AA} + {\bf M}_{AB} {\bf N}_{BA} & = & {\bf 1}
    \label{eq:partAA}, \\
    {\bf M}_{AA} {\bf N}_{AB} + {\bf M}_{AB} {\bf N}_{BB} & = & {\bf 0}
    \label{eq:partAB}, \\
    {\bf M}_{BA} {\bf N}_{AA} + {\bf M}_{BB} {\bf N}_{BA} & = & {\bf 0}
    \label{eq:partBA}, \\
    {\bf M}_{BB} {\bf N}_{BB} + {\bf M}_{BB} {\bf N}_{BB} & = & {\bf 1}
    \label{eq:partBB}.
  \end{eqnarray}
\end{mathletters}
From (\ref{eq:partBA}) we get
\begin{equation}
  {\bf N}_{BA} = - \bigl( {\bf M}_{BB} \bigr)^{-1} {\bf M}_{BA} {\bf
    N}_{AA}.
\end{equation}
When inserting in equation (\ref{eq:partAA}) and multiplying from the
right with $({\bf N}_{AA})$, we get
\begin{equation}
  {\bf M}_{AA} - {\bf M}_{AB} \bigl( {\bf M}_{BB} \bigr)^{-1} {\bf
    M}_{BA} =  \bigl( {\bf N}_{AA} \bigr)^{-1},
\end{equation}
and since $( {\bf N}_{AA})^{-1} = ( ({\bf M}^{-1})_{AA})^{-1}$,
equation (\ref{eq:tobeproven}) has been proven.

%\bibliographystyle{prsty}
%\bibliography{local}

\end{document}